\documentstyle[twoside,fleqn,espcrc2,epsf]{article}

\newcommand{\AmS}{{\protect\the\textfont2
  A\kern-.1667em\lower.5ex\hbox{M}\kern-.125emS}}

\title{The phase structure of the 3-d Thirring model}

\author{L. Del Debbio \address{Department of Physics, 
        University of Wales Swansea, 
        Swansea SA2 8PP, United Kingdom}%
        \thanks{This work is supported by an EC HMC Institutional 
                Fellowship under contract No. ERBCHBGCT930470.}
	(UKQCD Collaboration)}

\begin{document}

\begin{abstract}
We study the phase structure of the Thirring model in 3-d and find it
to be compatible with the existence of a non gaussian fixed point of
RG. A Finite Size Scaling argument is included in the equation of 
state in order to avoid the assumptions usually needed to extrapolate 
to the thermodymical limit.
\end{abstract}

\maketitle

\section{Introduction}

The Thirring model is a four-fermi theory with current-current interaction
defined by the following continuum Lagrangian:
\begin{equation}
{\cal L}=\bar\psi_i(\partial{\!\!\! /}\,+m)\psi_i
+{g^2\over{2N_f}}(\bar\psi_i\gamma_\mu\psi_i)^2
\end{equation}
where $\psi$ and $\bar\psi$ are four-fomponent spinors, $m$ is a bare mass, 
and $i$ runs over $N_f$ fermion species.

As in the case of the Gross-Neveu model (GN), na\"{\i}ve power counting 
suggests
that the theory is not renormalizable in the usual perturbative expansion
in powers of the coupling constant. Nonetheless, different approaches 
might allow a continuum limit to be defined for these theories. It has been 
established, using the $1/N_f$ expansion, that GN is renormalizable
to all orders, showing an interesting phase structure with an UV fixed
point of RG, spontaneous chiral symmetry breaking with dynamical mass
generation and a spectrum of mesonic bound states.

For the Thirring model the situation is less clear. The $1/N_f$ 
expansion turns out to be renormalizable (at least at leading order),
but neither chiral symmetry breaking nor charge renormalization are
observed (the $\beta$-function vanishes identically for all values
of the bare coupling).
On the other hand chiral symmetry breaking and dynamical mass generation
are expected at some critical value of the coupling $g_c(N_f)$, from the 
solution of Schwinger-Dyson equations. However,
in order to solve these equations, one has to use some Ansatz for
the 1-PI vertex function. According to the chosen Ansatz, the functional
dependence of $g_c$ on $N_f$ turns out to be different.
This inconsistent picture can be clarified by a lattice study of the model.

The existence of a non-gaussian fixed point, allowing a strongly
interacting theory with dynamical mass generation to be defined, 
would be of great theoretical
interest. It has a phenomenological relevance for model-building beyond
the Standard Model and it can describe high-$T_c$ superconductivity and
the IR behaviour of $QED_3$.
Besides, due to its relative simplicity, the 3-d model is an ideal 
laboratory to compare different non-perturbative approaches and to 
test algorithms for simulating dynamical fermions.

\section{Lattice formulation}

We simulate the theory using $N$ species of staggered fermions and 
a {\sl non-compact} action:
\begin{eqnarray}
S &=&\frac{1}{2} \sum \bar\chi_i(x) \eta_\mu(x) \left[ 1 + i A_\mu(x) \right]
\chi_i(x+\mu) \nonumber \\
&+&\mbox{h.c.}+ m \sum \bar\chi_i \chi_i + \frac{N}{4g^2} \sum A^{2}_{\mu} 
\end{eqnarray}
For $N=1$ this is equivalent to the {\sl compact} formulation where the
interaction term becomes $\bar\chi(x) \exp(iA_\mu(x)) \chi(x+\mu)$ and
the mass term for the vector field is no longer needed. 
Both lattice formulations have their drawbacks. For $N>1$, the compact version
generates higher order interactions (six fermions vertices) when the 
auxiliary field is integrated out. The non-compact one has an 
additive charge renormalization,
preventing to perform numerical simulations at arbitrary strong coupling
(see~\cite{ldd96} for details).

\section{Data analysis: the Equation of State}

The data are obtained, as usual, from numerical simulations at 
finite values of the 
bare mass and on finite lattices. In order to study dynamical mass 
generation, we need to consider the $m\rightarrow 0$ limit, while we also 
have to take into account the relevance of finite size effects. 
The only assumption we are going to make in analyzing our data is that
the RG flow for the Thirring model is similar to the one observed for
GN: we therefore expect to have a critical coupling $g_c(N)$, corresponding
to an UV fixed point, where a chiral symmetry breaking phase transition
occurs, separating a massless phase at weak coupling from a massive one 
at strong coupling. Being below the upper critical dimensions ($d=4$),
one can solve the RGE without worrying about logarithmic corrections to
scaling. 

An equation of state can then be derived using the solution of the RG 
equations for the external magnetic field at given 
magnetization~\cite{zinn93}. In our
case, the external, symmetry-breaking field, is the bare mass $m$, and
the magnetization is the order parameter monitoring the symmetry 
breaking, i.e. the chiral condensate. In terms of the scaled variables, we
obtain:
\begin{equation}
m \langle\bar\psi\psi\rangle^{-\delta} = {\cal F}_1\left( 
t \langle\bar\psi\psi\rangle^{-1/\beta}\right)
\end{equation}
where $t=1/g^2-1/g_{c}^{2}$ and ${\cal F}_1$ is a universal scaling function.

In order to take into account finite size effects, we introduce the 
volume of the lattice as a scaling field of dimension $d$. The scaling
function depends now on two independent combinations of rescaled 
variables:
\begin{equation}
\label{eq:eos}
m \langle\bar\psi\psi\rangle^{-\delta} = {\cal F}_2\left( 
t \langle\bar\psi\psi\rangle^{-1/\beta}, L^{-1/\nu} 
\langle\bar\psi\psi\rangle^{-1/\beta} \right)
\end{equation}
where $L$ is the linear size of the lattice.

\begin{figure}[ht]
\vspace{3pt}
\centerline{
\setlength\epsfxsize{200pt}\epsfbox{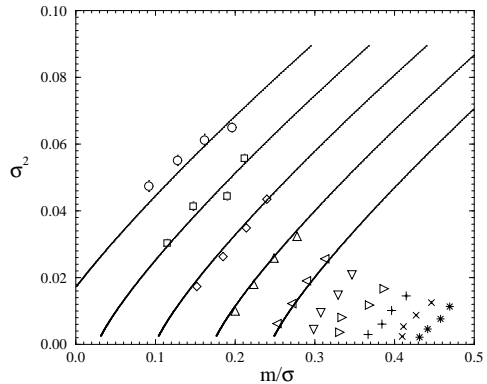}}
\caption{Fisher plot for $N=1$, from data at $\beta=1.6 (\triangle)$,
$1.8 (\triangleleft)$, $2.0 (\nabla)$, $2.2 (\triangleright)$, $2.4 (+)$,
$2.6 (\times)$, on a $12^3$ lattice. \label{fig:fisher}}
\end{figure}

In the limits $L\rightarrow\infty$, $t=0$ or $m=0$, Eq.~(\ref{eq:eos}) 
reproduces the usual definitions of the critical exponents $\nu$,
$\delta$ and $\beta$. Neglecting the finite-size correction and performing
a Taylor expansion of ${\cal F}_1$ for small $t$, one obtains a five 
parameter equation
to which the lattice data at fixed lattice volume can be fitted, allowing
the critical exponents introduced above to be determined:
\begin{equation}
\label{eq:fit1}
m = B \langle\bar\psi\psi\rangle + A t \langle\bar\psi\psi\rangle^{\delta
-1/\beta}
\end{equation}
Our results for $N=1,2$ are reported in Tab~(\ref{tab:fit}), 
where Fit I corresponds to Eq.~(\ref{eq:fit1}), while for Fit II we set
$\delta-1/\beta=1$, as suggested in~\cite{dagotto90}.
It is intersting to remark that, within errors, the hypothesis underlying
Fit II is compatible with the result obtained with $\beta$ as a free 
parameter. Since Eq.~(\ref{eq:fit1}) is obtained from a Taylor expansion,
we expect it to be valid only for a limited range of $t$ values. The number
of points included in our fits has been chosen in order to minimize the
$\chi^2$/dof. The good quality of the fit can be seen as an evidence
in favour of the description of the phase structure of the Thirring model
in terms of a non-gaussian, UV stable, fixed point. Moreover, the critical 
exponents are significantly different from their mean-field values; 
i.e. the continuum theory defined by the lattice model is not trivial.

\begin{table*}[hbt]
\setlength{\tabcolsep}{1.5pc}
\newlength{\digitwidth} \settowidth{\digitwidth}{\rm 0}
\catcode`?=\active \def?{\kern\digitwidth}
\caption{Results from fit}
\label{tab:fit}
\begin{tabular*}{\textwidth}{@{}l@{\extracolsep{\fill}}rrrrr}
\hline
	& Parameter & Fit I & Fit II & Fit III &Mean Field \\
\hline
$N=1$	& $1/g_{c}^{2}$ & $2.03(9)$  & $1.94(4)$  & $2.03(3)$  & --    \\
	& $\delta$      & $2.32(23)$ & $2.68(16)$ & $2.65(7)$  & $3$   \\
	& $\beta$       & $0.71(9)$  &    --      &    --      & $1/2$ \\
        & $\nu$         &    --      &    --      & $0.80(8)$  & $1/2$ \\
	& $A$           & $0.32(5)$  & $0.37(1)$  & $0.33(9)$  & --    \\
	& $B$           & $1.92(43)$ & $2.86(53)$ & $2.33(19)$ & --    \\
        & $\chi^2$/dof  & $2.4$      & $2.1$      & $1.5$      &       \\
\\
$N=2$	& $1/g_{c}^{2}$ & $0.63(1)$  & $0.66(1)$  &   --       & --    \\
	& $\delta$      & $3.67(28)$ & $3.43(19)$ &   --       & $3$   \\
	& $\beta$       & $0.38(4)$  &    --      &   --       & $1/2$ \\
	& $A$           & $0.78(5)$  & $0.73(2)$  &   --       & --    \\
	& $B$           & $7.9(2.8)$ & $6.4(1.5)$ &   --       & --    \\
        & $\chi^2$/dof  & $3.1$      & $2.0$      &   --       &       \\
\hline
\end{tabular*}
\end{table*}
If the critical behaviour is described by mean-field theory, the
equation of state becomes:
\begin{equation}
\langle\bar\psi\psi\rangle^2 = \kappa_1 {m \over \langle\bar\psi\psi\rangle}
+ \kappa_2 t
\end{equation}
so that $\langle\bar\psi\psi\rangle^2$ is a linear function of the ratio
$m/\langle\bar\psi\psi\rangle$ and a positive value of the intercept 
corresponds to a non-vanshing value of the chiral condensate for $m=0$, 
while the intercept will be exactly zero at the critical coupling. 
This plot, known as Fisher plot, is shown in Fig.~(\ref{fig:fisher}) for
$N=1$, where a clear evidence for chiral symmetry breaking can be seen.

Finally, one can include finite size effects into the analysis, by 
expanding ${\cal F}_2$ in Eq.~(\ref{eq:eos}). After setting 
$\delta-1/\beta=1$ in order to reduce the number of parameters, 
an equation to fit data from different lattice sizes can be 
written as:
\begin{equation}
m = A \langle\bar\psi\psi\rangle^\delta + B (t + C L^{-1/\nu})
\langle\bar\psi\psi\rangle
\end{equation}
The results of the fit for $N=1$ are reported in Tab.~(\ref{tab:fit}) as
Fit III. They include data for different values of $t$ and $m$, from
$8^3$, $12^3$ and $16^3$ lattices.
The critical exponent $\eta$ is defined in terms of $\delta$ and the 
space-time dimension $d$ using hyperscaling, 
while $\beta$ has been fixed by the above 
constraint. Using the results from Fit III, $\beta$, $\nu$ and $\eta$
verify, within errors, the scaling relation:
\begin{equation}
\beta = \frac{1}{2} \nu (d-2+\eta)
\end{equation}
confirming the fact that our data can be described by a UV fixed point 
in the RG flow of the model.

\section{Conclusions}

Our results can be summarized as follows. We have found a numerical 
evidence for a chiral symmetry breaking phase transition in the 
Thirring model with $N=1,2$, showing a breakdown of the leading
order $1/N$ prediction for small $N$. A more careful analysis
of the $N=3$ case could also shed some light on the different scenarios
predicted by the Schwinger-Dyson approach.  Our data are consistent with
the existence of a non-gaussian fixed point, corresponding to this phase
transition. By combining the equation of state with finite size scaling 
analysis, we are able to determine the critical exponents. These indicate
that the continuum theories defined at the critical points are not 
mean-field. 
In order to clarify this structure, we are now studying the spectrum of
the theory in the different phases, the renormalized coupling and
the lines of constant physics. 
A detailed description of our results
will appear soon~\cite{ldd96b}.

\end{document}